# Orbital-selective Mottness Driven by Geometric Frustration of Interorbital Hybridization in Pr$_4$Ni$_3$O$_{10}$

Yidian Li[1*], Mingxin Zhang[2,3*], Xian Du[4], Cuiying Pei[2], Jieyi Liu[5], Houke Chen[6], Wenxuan Zhao[1], Kaiyi Zhai[1], Yinqi Hu[1], Senyao Zhang[1], Jiawei Shao[1], Mingxin Mao[1], Yantao Cao[7], Jinkui Zhao[3], Zhengtai Li[8], Dawei Shen[9], Yaobo Huang[8], Makoto Hashimoto[10], Donghui Lu[10], Zhongkai Liu[2,11], Yulin Chen[2,6,11], Hanjie Guo[3], Yilin Wang[12,13†], Yanpeng Qi[2,11,14†], and Lexian Yang[1,15†]

[1]*State Key Laboratory of Low Dimensional Quantum Physics, Department of Physics, Tsinghua University, Beijing 100084, China*

[2]*State Key Laboratory of Quantum Functional Materials, School of Physical Science and Technology, ShanghaiTech University, Shanghai 201210, China*

[3]*Songshan Lake Materials Laboratory, Dongguan, Guangdong 523808, China*

[4]*Department of Applied Physics, Yale University, New Haven, CT 06511, USA*

[5]*Diamond Light Source, Harwell Science and Innovation Campus, Didcot OX11 0DE, UK*

[6]*Department of Physics, Clarendon Laboratory, University of Oxford, Parks Road, Oxford OX1 3PU, UK*

[7]*Key Lab for Magnetism and Magnetic Materials of the Ministry of Education, Lanzhou University, Lanzhou 730000, China*

[8]*Shanghai Synchrotron Radiation Facility, Shanghai Advanced Research Institute, Chinese Academy of Sciences, Shanghai 201210, China*

[9]*School of Nuclear Science and Technology and National Synchrotron Radiation Laboratory, University of Science and Technology of China, Hefei, Anhui 230029, China*

[10]*Stanford Synchrotron Radiation Lightsource, SLAC National Accelerator Laboratory, Menlo Park, CA 94025, USA*

[11]*ShanghaiTech Laboratory for Topological Physics, Shanghai 200031, China*

[12]*School of Emerging Technology, University of Science and Technology of China, Hefei 230026, China*

[13]*Hefei National Laboratory, University of Science and Technology of China, Hefei 230088, China*

[14]*Shanghai Key Laboratory of High-resolution Electron Microscopy, ShanghaiTech University, Shanghai 201210, China*

[15]*Frontier Science Center for Quantum Information, Beijing 100084, China*

*These authors contributed equally to this work*

*Emails: LXY: lxyang@tsinghua.edu.cn; YPQ: qiyp@shanghaitech.edu.cn; YLW: yilinwang@ustc.edu.cn;*

The interplay among orbital-selective Mott physics, Hund's coupling, tunable structural motifs, and Kondo-like scattering establishes a compelling paradigm for understanding and engineering correlated multi-orbital systems, as vividly exemplified by nickelate superconductors. Here, using high-resolution angle-resolved photoemission spectroscopy combined with theoretical calculations, we systematically investigate the electronic properties of trilayer nickelates. In $La_4Ni_3O_{10}$, we observe pronounced interorbital hybridization, whereas in $Pr_4Ni_3O_{10}$, the flat $d_{z^2}$ band becomes markedly incoherent and diminishes in spectral weight. By contrast, the dispersive $d_{x^2-y^2}$ bands retain coherence in both compounds. This striking incoherence/coherence dichotomy identifies an orbital-selective Mott phase modulated by the interlayer Ni-O-Ni bonding angle. The depletion of the $d_{z^2}$ orbitals further frustrates the interorbital hybridization and influences the density-wave transition in $Pr_4Ni_3O_{10}$. Moreover, the density-wave gap is substantially reduced in $Pr_4Ni_3O_{10}$, likely due to extra scattering channels provided by the local moments of $Pr^{3+}$ cations. Our findings elucidate the intricate interplay among lattice, orbital, spin, and electronic degrees of freedom and reveal a feasible structural control parameter for the multi-orbital correlated state in trilayer nickelates, which provide a concrete framework for understanding the emergence of superconductivity under high pressure.

Layered nickelates in the Ruddlesden-Popper (R-P) phase have emerged as a compelling platform for exploring correlated superconductivity in 3$d$ transition metal oxides [1-10]. The superconductivity in these systems emerges from a delicate balance among different microscopic interactions, including orbital-dependent electron correlation, interorbital hybridization modulated by Hund's coupling ($J_H$), electron-phonon interaction, and interlayer correlation [11-28]. This intricate interplay also creates an intriguing landscape of emergent phenomena, such as charge/spin-density-waves [29-32], the strange metal phase, and non-Fermi liquid behaviors [33,34], which collectively challenge conventional paradigms of superconductivity. To decipher the key factors that govern the physics of nickelates, it is essential to map their electronic structures across the phase diagram.

In contrast to the bilayer nickelates with structural complexity and imperfections [13,35-37], trilayer R-P phase nickelates, though typically exhibiting lower superconducting transition temperature ($T_c$), offer a more pristine platform for investigating the electronic structures, microscopic interactions, and competing density-wave orders [38,39]. In particular, $Pr_4Ni_3O_{10}$ was recently discovered to be superconducting with an onset $T_c \approx 40$ K under pressure [8,40,41], significantly outperforming its sibling compound $La_4Ni_3O_{10}$. The density-wave transition temperature ($T_{dw}$) is also elevated in $Pr_4Ni_3O_{10}$, suggesting a subtle correlation between $T_c$ and $T_{dw}$ [42], Apparently, the rare-earth element substitution plays an essential role in tuning the physical properties of nickelates, yet its impact on their electronic structures remains experimentally elusive. The $Pr^{3+}$ cations introduce dual effects: the smaller ionic radius imposes effective chemical pressure, while their local moments act as Kondo-like scattering centers [43]. Unraveling how the orbital-selective correlation, Hund's coupling, tunable structural motifs, and potential Kondo-like scattering cooperate and/or compete in the trilayer nickelates is crucial for understanding the normal-state density-waves and the enhanced superconductivity in $Pr_4Ni_3O_{10}$.

In this work, by combining high-resolution angle-resolved photoemission spectroscopy (ARPES) measurements and theoretical calculations, we systematically investigate the low-energy electronic structures of $Pr_4Ni_3O_{10}$ and $La_4Ni_3O_{10}$ single crystals. The two trilayer nickelates exhibit an overall similar orbital-dependent electron correlation effect. In $La_4Ni_3O_{10}$, a prominent flat band derived from $d_{z^2}$ orbitals resides approximately 30 meV below $E_F$, hybridizing with the dispersive $d_{x^2-y^2}$ band. In contrast, in $Pr_4Ni_3O_{10}$, the $d_{z^2}$ band becomes significantly incoherent with depleted spectral weight, while the $d_{x^2-y^2}$ band remains coherent. Our dynamical-mean-field-theory (DMFT) calculations well reproduce the observed coherence/incoherence dichotomy by tuning the interlayer Ni-O-Ni bonding angle and Hund's coupling. We propose that the site-selective electron correlation frustrates the interorbital hybridization and drives $Pr_4Ni_3O_{10}$ into an orbital-selective Mott (OSM) phase. Moreover, while the enhanced electron correlation effect of the $d_{z^2}$ flat band near the $E_F$ favors the density-wave orders, the density-wave gap is significantly reduced in $Pr_4Ni_3O_{10}$, likely due to the perturbation from the local moments of $Pr^{3+}$ cations. The revelation of the structural tunable OSM phase establishes a versatile landscape to explore and engineer the correlated physics, including the normal-state density-wave orders and the superconductivity emerging under high pressure.

Figures 1(a)–(c) compare the crystal structures of $Pr_4Ni_3O_{10}$ and $La_4Ni_3O_{10}$. Both compounds crystallize into the same space group, with Pr/La atoms occupying equivalent atomic positions [40,44-47]. In each unit cell, there are three Ni-O layers forming a trilayer building block. At ambient pressure, the distortion of $NiO_6$ octahedra along the *c*-axis doubles the unit cell within the *ab*-plane. The smaller ionic radius of $Pr^{3+}$ (compared to $La^{3+}$) reduces the Ni-O-Ni bonding angle and, consequently, shortens the interlayer

spacing, applying effective chemical pressure that is suggested to be crucial for the enhanced superconductivity in $Pr_4Ni_3O_{10}$ [8]. The substitution of La with Pr further amplifies the structural distortion at ambient pressure: the interlayer Ni-O-Ni bonding angle reduces from 165° to 158° [Figs. 1(b)–1(c)]. The interlayer bonding angle provides a delicate structural degree of freedom for engineering the electronic properties of trilayer nickelates.

Single crystal X-ray diffraction measurement confirms the crystal structure of our $Pr_4Ni_3O_{10}$ samples in Fig. 1(d). The temperature-dependent resistivity of $Pr_4Ni_3O_{10}$ and $La_4Ni_3O_{10}$ shows an upturn at about 156 K and 135 K, respectively [Fig. 1(e)], which were attributed to the intertwined charge/spin-density-wave transitions [48,49]. The magnetic susceptibility of $Pr_4Ni_3O_{10}$ gradually increases with decreasing temperature, following a prototypical Curie-Weiss paramagnetic behavior [Fig. 1(f)]. A noticeable anomaly at $T_{dw}$ in the derivative of the susceptibility [the inset of Fig. 1(f)] suggests an intimate connection between the magnetism and the density-wave transition. On the drastic contrary, the magnetic susceptibility of $La_4Ni_3O_{10}$ first decreases at high temperatures, then increases below $T_{dw}$ [43,44,46,50,51]. Apparently, the large paramagnetic moments of $Pr^{3+}$ cations mask the intrinsic magnetism of the Ni-O layers, which also influence the electronic properties of $Pr_4Ni_3O_{10}$ [43,44,52-54].

Next, we compare the basic electronic structures of $Pr_4Ni_3O_{10}$ and $La_4Ni_3O_{10}$ using high-resolution ARPES. Figures 2(a)–2(c) show the Fermi surface obtained by integrating the photoemission intensity within a 40 meV energy window around $E_F$. We observe large portions of straight Fermi surface segments (β pocket), in agreement with previous results and density-functional-theory (DFT) calculations [38,55-57]. The α pocket appears more elliptical in $Pr_4Ni_3O_{10}$, reflecting a larger orthorhombic anisotropy (i.e., larger difference between the in-plane lattice constants *a* and *b*) compared to $La_4Ni_3O_{10}$. As indicated by the blue and red arrows in Figs. 2(a) and 2(c), we resolve clear folding of the β pocket associated with the density-wave transition. The corresponding wave vector $q_{dw}$ is about $0.62b^*$, consistent with previous scattering and scanning tunneling microscopy experiments [48,58].

Figures 2(d)–2(e) compare the band dispersions along high-symmetry directions. In $La_4Ni_3O_{10}$, dispersive $d_{x^2-y^2}$ bands crossing $E_F$ form the α and β Fermi pockets. The $d_{z^2}$ orbital, on the other hand, contributes to a flat band γ at about 30 meV below $E_F$ near the $\bar{\Gamma}'$ point. Prominently, we resolve clear hybridization between the $d_{x^2-y^2}$ and $d_{z^2}$ bands [Fig. 2(g) and the Supplemental Material [59]]. The interorbital hybridization tends to delocalize electrons between the $d_{x^2-y^2}$ and $d_{z^2}$ orbitals and competes with the Hund's coupling that favors localized, high-spin moments. This intricate interplay plays a foundational role in the physics of nickelates, including normal-state magnetic and density-wave properties. In particular, the interorbital hybridization is expected to be significantly enhanced under high pressure, directly influencing the superconducting phase diagram [20,22,60-65].

Overall, the band structure of $Pr_4Ni_3O_{10}$ is closely analogous to that of $La_4Ni_3O_{10}$. However, the $d_{z^2}$ flat band of $Pr_4Ni_3O_{10}$ is significantly incoherent with depleted spectral weight near $E_F$ [Figs. 2(d) and 2(f)]. It is almost undetectable at relatively low photon energies, with only a weak, incoherent spectral feature close to $E_F$ resolvable at high photon energies [Fig. 2(f)], compared to the prominent flat $d_{z^2}$ band of $La_4Ni_3O_{10}$ [Figs. 2(e) and 2(g)]. We emphasize that the dispersive $d_{x^2-y^2}$ bands remain clearly visible in $Pr_4Ni_3O_{10}$ [Fig. 2(d)], with spectral width comparable to those of $La_4Ni_3O_{10}$ [Fig. 2(e) and the Supplemental Material [59]]. By comparing the experimental and DFT-calculated band structures, we estimate the band renormalization factors to be about 3 and 6 for the $d_{x^2-y^2}$ and $d_{z^2}$ bands, similar to the values in $La_4Ni_3O_{10}$ [38]. This orbital-dependent band renormalization and the diminishing $d_{z^2}$ bands identify an OSM phase in $Pr_4Ni_3O_{10}$, highlighting the similarity between the multi-orbital physics in iron-based superconductors and nickelates [66].

The diminishing and incoherent nature of the $d_{z^2}$ bands in $Pr_4Ni_3O_{10}$ is counterintuitive, as the reduced interlayer spacing would typically enhance interlayer hopping and quasiparticle coherence. To understand this non-trivial observation, we performed systematic DFT+DMFT calculations in Figs. 3(a)–3(e). The Hund's coupling $J_H = 0.8$ eV and on-site Coulomb interaction $U = 5$ eV were chosen to match the ARPES experiments. Layer-selective calculations suggest that the flat bands near $E_F$ originate primarily from the $d_{z^2}$ orbitals in the outer Ni-O layers [Figs. 3(a)–3(b)]. Our DFT+DMFT calculations successfully reproduce the key experimental findings: $La_4Ni_3O_{10}$ exhibits coherent $d_{x^2-y^2}$ and $d_{z^2}$ bands [Fig. 3(c)], in drastic contrast to the orbital-selective coherence-incoherence dichotomy between these two orbitals in $Pr_4Ni_3O_{10}$ [Fig. 3(d)]. Particularly, the $d_{z^2}$ bands are diminished and almost unresolvable in $Pr_4Ni_3O_{10}$, in good agreement with the experiment. Interestingly, by replacing Pr with La while maintaining the structural parameters of $Pr_4Ni_3O_{10}$, the spectral function of this hypothetical $La_4Ni_3O_{10}$ closely resembles that of $Pr_4Ni_3O_{10}$ [Fig. 3(e)]. Therefore, the coherence of the $d_{z^2}$ bands is vividly modulated by the interlayer Ni-O-Ni bonding angle, establishing it as a feasible structural parameter for controlling the OSM phase in $Pr_4Ni_3O_{10}$.

Notably, the calculations also suggest an enhanced renormalization of $d_{x^2-y^2}$ band by reducing the interlayer Ni-O-Ni bonding angle. To confirm this observation, we performed laser-ARPES measurements in Figs. 3(f)–3(h). In good agreement with the calculations in Figs. 3(c)–3(d), the $d_{x^2-y^2}$ band exhibits a correlation-induced dispersion anomaly ("waterfall"-like dispersion) at a lower energy in $Pr_4Ni_3O_{10}$ than in $La_4Ni_3O_{10}$. This result is consistent with the incoherence of the $d_{z^2}$ band, which frustrates the interorbital hybridization and, in turn, eliminates a vital electron delocalization channel for the $d_{x^2-y^2}$ band. Moreover, it reduces the screening from the $d_{z^2}$ electrons. Consequently, the correlation effect of the $d_{x^2-y^2}$ bands is enhanced.

After revealing the similarities and differences between the electronic properties of the two compounds, we next examine their density-wave transitions. As already shown in Figs. 2(a)–2(c), both systems exhibit clear band folding associated with the density-wave transition, with similar wave vectors. To further quantify the density-wave gap, we performed high-resolution laser-ARPES experiments in Fig. 4. In La$_4$Ni$_3$O$_{10}$, we observe similar band renormalization along $\bar{\Gamma}\bar{X}$ and $\bar{\Gamma}\bar{M}$ [Fig. 4(a)]. The α/β splitting is clearly resolved along $\bar{\Gamma}\bar{X}$. At the Fermi momenta ($k_F$), the β band exhibits a leading-edge shift of about 6 meV with the temperature, while the α band remains nearly gapless over a large temperature range [Fig. 4(b)]. Interestingly, a pronounced leading-edge shift occurs at $k_F^\alpha$ below $T_{\text{dw}}$ along $\bar{\Gamma}\bar{M}$ [Figs. 4(c)–4(d)], suggesting an anisotropy of the density-wave gap. A Bardeen-Cooper-Schrieffer (BCS)-type analysis of this shift yields a zero-temperature leading-edge gap size ($\Delta_0$) of approximately 12 meV [Fig. 4(d)]. A similar anisotropic density-wave gap is resolved in Pr$_4$Ni$_3$O$_{10}$ [Fig. 4(e)]. Nevertheless, despite its higher $T_{\text{dw}}$, Pr$_4$Ni$_3$O$_{10}$ exhibits a significantly smaller leading-edge gap ($\Delta_0 \approx 6$ meV). The corresponding reduced gap value $2\Delta_0/k_B T_{dw}$ is estimated to be about 0.9 and 2.1 for Pr$_4$Ni$_3$O$_{10}$ and La$_4$Ni$_3$O$_{10}$, respectively.

The different density-wave behaviors in the trilayer nickelates can be understood by the orbital-selective electron correlation. On the one hand, the enhanced correlation of $d_{z^2}$ orbitals favors the magnetic instability, which can likely elevate the $T_{\text{dw}}$. On the other hand, the local magnetic moment of Pr$^{3+}$ cations may act as Kondo-like scattering centers, resonating with the orbital-selective correlation effect and further driving the $d_{z^2}$ orbitals into incoherent states. Since the magnetic $Pr^{3+}$ cations reside in the rocksalt-type spacing layer between trilayer blocks [Fig. 1(a)], they primarily affect the $d_{z^2}$ orbital in the outer Ni-O layers. Furthermore, the $Pr^{3+}$ local moments create additional spin fluctuation channels and local magnetic fields that may disrupt the phase stiffness of long-range density-wave order and reduce the gap in $Pr_4Ni_3O_{10}$.

In summary, the strong interorbital hybridization that is modulated by Hund's coupling, electron correlations, and interlayer interactions renders the intertwined density-waves in the normal state of $La_4Ni_3O_{10}$, from which the superconductivity emerges under high pressure. By contrast, the interorbital hybridization is frustrated in $Pr_4Ni_3O_{10}$, establishing an OSM phase that is tunable through a geometric parameter. Our results suggest an intriguing quantum critical point between the OSM phase and pressurized superconductivity in $Pr_4Ni_3O_{10}$, which can be tuned through the interlayer bonding angle. Furthermore, our work highlights the similarity between the multi-orbital superconductivity in nickelates and iron-based superconductors.

*Acknowledgements*—We thank J. Osiecki and C. M. Polley of MAX IV for supporting the beamtime. This work is funded by the National Key R&D Program of China (Grant No. 2022YFA1403100 and No. 2022YFA1403201), the National Natural Science Foundation of China (No. 92365204 and No. 12274251). Y.P.Q. acknowledges support from the Science and Technology Commission of Shanghai Municipality (Grant No. 25DZ3008200). Y.L.W. acknowledges support from the Quantum Science and Technology-National Science and Technology Major Project (No. 2021ZD0302803) and the National Key R&D Program of China (Grant No. 2023YFA1406304).

*Data availability*—The data sets that support the findings of this study are available from the corresponding author upon request.

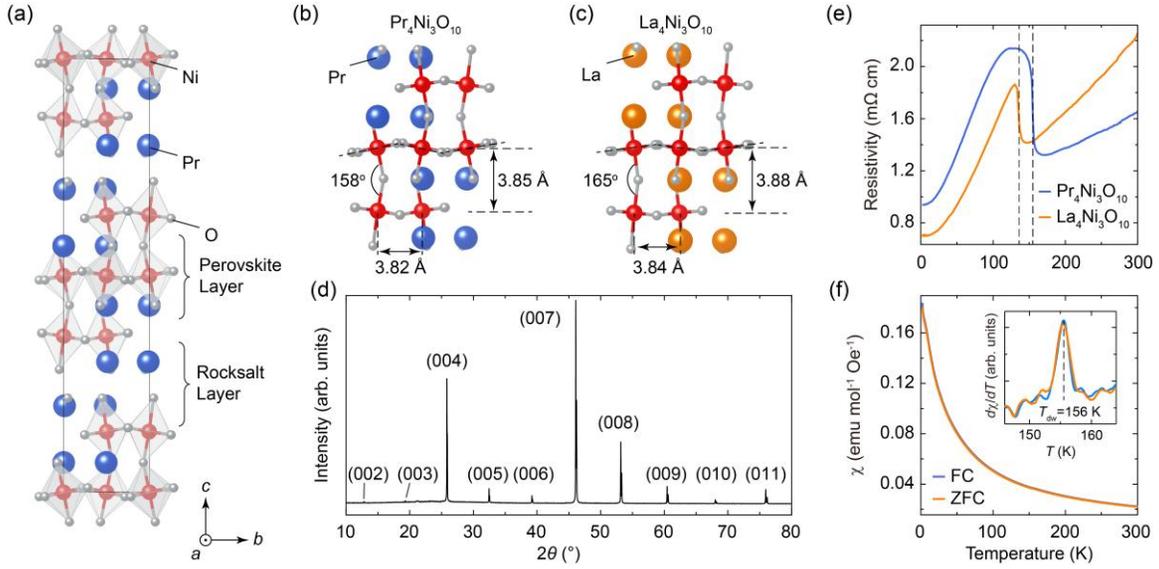

Fig. 1. Crystal structure and transport properties of $Pr_4Ni_3O_{10}$. (a) Schematic three-dimensional structure of $Pr_4Ni_3O_{10}$ single crystal. (b)–(c) Comparison of the crystal structures of (b) $Pr_4Ni_3O_{10}$ and (c) $La_4Ni_3O_{10}$ along the *a*-axis. The interlayer Ni-O-Ni bonding angles are indicated. (d) Single crystal X-ray diffraction of $Pr_4Ni_3O_{10}$. (e) Temperature-dependent resistivity of $Pr_4Ni_3O_{10}$ compared with $La_4Ni_3O_{10}$ at ambient pressure. The dashed lines indicate the density-wave transition temperatures ($T_{dw}$). (f) Magnetic susceptibility ($\chi$) of $Pr_4Ni_3O_{10}$ under field-cooling (FC) and zero-field-cooling (ZFC) conditions. The inset shows the zoom-in plot of the derivative of the susceptibility with respect to the temperature showing the density-wave transition around 156 K.

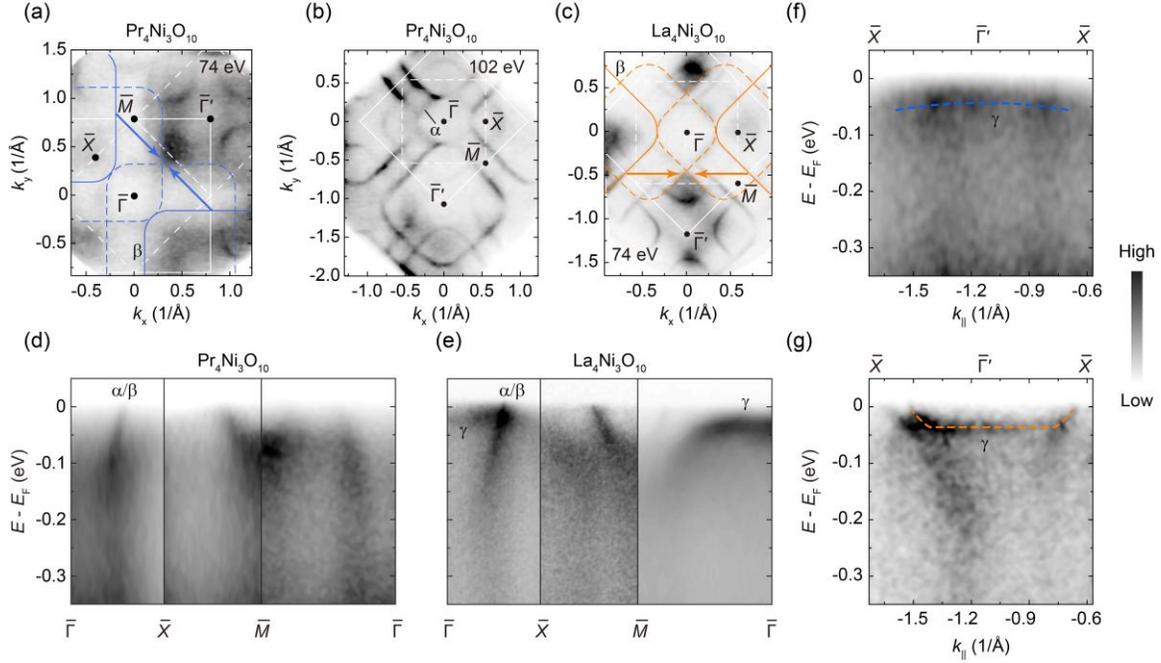

Fig. 2. Basic electronic structure of trilayer nickelates. (a)–(c) Fermi surface maps of (a)–(b) Pr$_4$Ni$_3$O$_{10}$, and (c) La$_4$Ni$_3$O$_{10}$. The solid and dashed white lines indicate the Brillouin zones (BZs) based on the primitive and conventional unit cells, respectively. The solid blue and orange lines are the guides to the eyes of the β pockets. The dashed blue and orange lines represent the folded Fermi sheets, connected by the density-wave vectors $q_{dw}$ (blue and orange arrows). (d)–(e) Band dispersions of (d) Pr$_4$Ni$_3$O$_{10}$ and (e) La$_4$Ni$_3$O$_{10}$ along high symmetry directions. (f)–(g) Comparison between the flat $d_{z^2}$ band (γ) of (f) Pr$_4$Ni$_3$O$_{10}$ and (g) La$_4$Ni$_3$O$_{10}$. Data in (d) and (e) were collected using 102 eV and 74 eV photons. Data in (f) and (g) were collected using 160 eV and 104 eV photons. All data were collected below 22 K. More data on the γ bands are provided in the Supplemental Material [59].

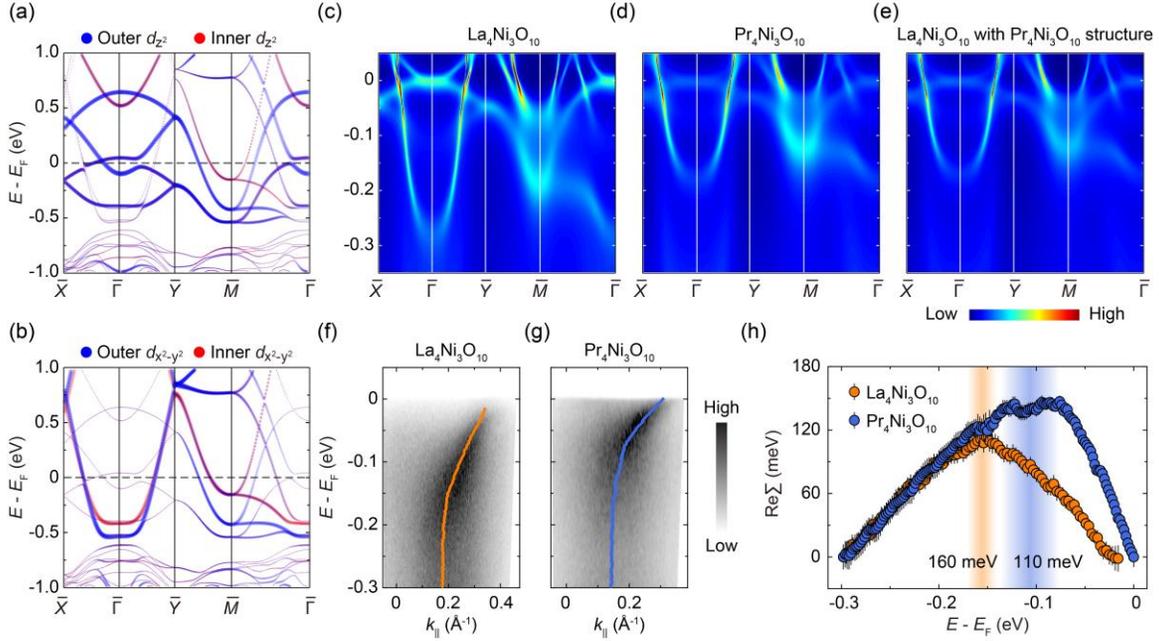

Fig. 3. Geometrically tunable electron correlations in trilayer nickelates. (a)–(b) the projected Ni-layer-dependent orbital characters of (a) the $d_{z^2}$ bands and (b) the $d_{x^2-y^2}$ bands of Pr$_4$Ni$_3$O$_{10}$. (c)–(d) Density-functional-theory (DFT) + Dynamical-mean-field-theory (DMFT) calculations of the spectral function of (c) La$_4$Ni$_3$O$_{10}$ and (d) Pr$_4$Ni$_3$O$_{10}$. (e) Calculated spectral function of La$_4$Ni$_3$O$_{10}$ by artificially replacing Pr with La atoms while maintaining the structural parameters of Pr$_4$Ni$_3$O$_{10}$. The calculations in (c)–(e) were performed with Hund's coupling $J_H$ = 0.8 eV and on-site Coulomb interaction $U$ = 5 eV at 100 K. (f)–(g) band dispersion of (f) La$_4$Ni$_3$O$_{10}$ and (g) Pr$_4$Ni$_3$O$_{10}$ along the $\bar{\Gamma}\bar{M}$ direction measured with laser-ARPES. The orange and blue curves are the extracted band dispersions by fitting the momentum-distribution curves to Lorentzians. (h) The real-part of electron self-energies (Re$\Sigma$), showing dispersion anomalies at about 110 meV and 160 meV in Pr$_4$Ni$_3$O$_{10}$ and La$_4$Ni$_3$O$_{10}$, respectively.

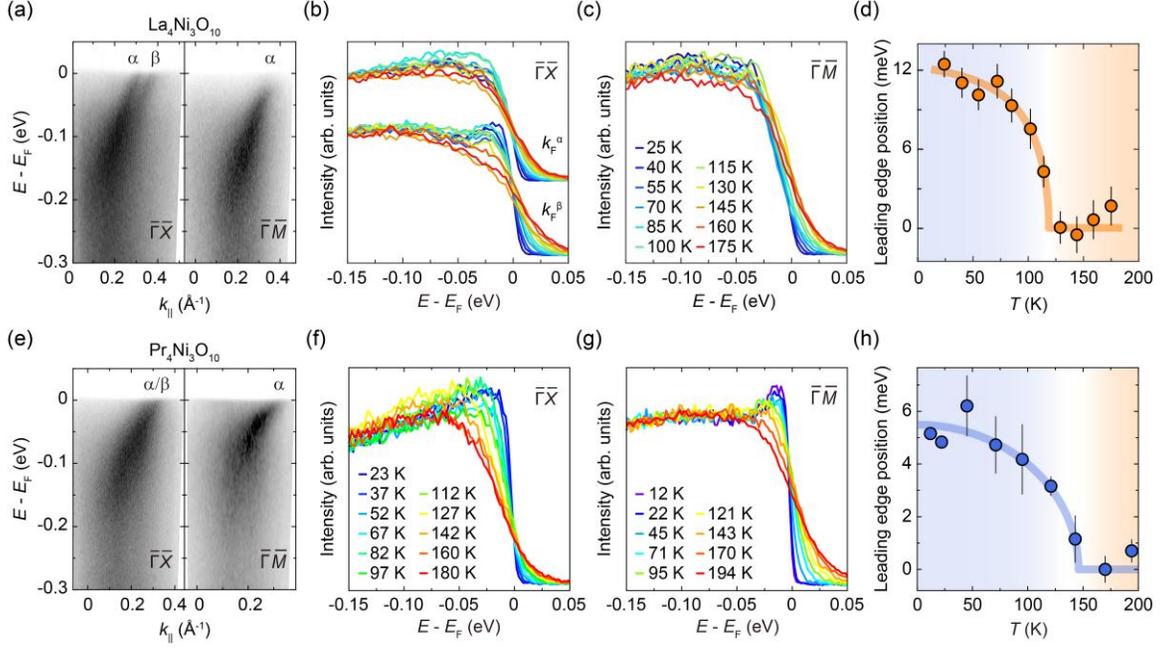

Fig. 4. Anisotropic density-wave gaps of trilayer R-P phase nickelates. (a) Band dispersions of $La_4Ni_3O_{10}$ along the $\overline{\Gamma}\overline{X}$ and $\overline{\Gamma}\overline{M}$ directions. (b)–(c) Energy-distribution curves (EDCs) at the Fermi momenta ($k_F$) along the (b) $\overline{\Gamma}\overline{X}$ and (c) $\overline{\Gamma}\overline{M}$ directions at selected temperatures. (d) Temperature-dependence of the leading-edge positions of EDCs along the $\overline{\Gamma}\overline{M}$ direction. (e)–(h) The same as (a)–(d) but for $Pr_4Ni_3O_{10}$. All data were collected using 7 eV laser.

Crossover to an Orbital-Selective Mott Phase in $A_x$Fe$_{2-y}$Se$_2$ ($A$ = K, Rb) Superconductors, Phys. Rev. Lett. **110**, 067003 (2013).